# A Rule-Based Short Query Intent Identification System

Arijit De[1], Sunil Kumar Kopparapu[2]

*TCS Innovation Labs-Mumbai*

*Tata Consultancy Services*

*Pokhran Road No. 2, Thane West, Maharashtra 400601, India*

[1]`arijit6.d@tcs.com`

[2]`sunilkumar.kopparapu@tcs.com`

*Abstract*—Using SMS (Short Message System), cell phones can be used to query for information about various topics. In an SMS based search system, one of the key problems is to identify a domain (broad topic) associated with the user query; so that a more comprehensive search can be carried out by the domain specific search engine. In this paper we use a rule based approach, to identify the domain, called Short Query Intent Identification System (SQIIS). We construct two different rule-bases using different strategies to suit query intent identification. We evaluate the two rule-bases experimentally.

## I. INTRODUCTION

In a typical Information Retrieval (IR) system, the user enters a query in the form of a sentence or a string of words. The task of the system is to match the query to related documents in a rank ordered list of varying relevance. The performance of retrieval can be improved through identification of the *intent* of the query; the identified intent can be used to enable a specialized and hence restricted domain search. While, query intent identification becomes more important when topically different data repositories are being searched through a single interface, within a metasearch like framework, it becomes a crucial for SMS based mobile search. SMS based domain specific question answering systems are beginning to get popular [8, 9, 12]. Suktarachan [12] proposes such a system for answering questions of farmers while [9] proposes for FAQ and [8] for yellow pages query. However, a mobile phone user's queries might vary from sports scores, movies currently playing, yellow pages to find the nearest restaurant, route directions to a particular even to give a few examples. Almost always the answer to these queries will need searching distinctly different data repositories or domain specific systems. In such case, the first step in the search process is to correctly identify which domain (broad topic) the query is related to and enable a corresponding data repository where the specific search for information be carried out.

One luxury that an Internet search system has over a SMS based mobile search system is that the former can return a list of documents that it thinks fit the query of the user. The user can then pick and chose what is relevant. However, the answer to an SMS query has to be short and of course accurate. Thus, SMS based systems can not dump a whole load of information to the user. This essentially means accuracy is of paramount importance in SMS searches.

In this paper, we describe the working of Short Query Intent Identification System (SQIIS), which can be used to accurately detect the domain of an SMS query. The key strength of our system is a set of rules (rule-based) which enables the system identify the intent of the query and hence the domain. Our focus is in analyzing the construction of two different rule bases that can be used in domain detection.

The paper is organized as follows. In Section II we give a brief overview of previous work in query intent identification and discuss how our work is related and different from the general body of knowledge in this area. In Section III we provide a brief introduction to SQIIS followed by a discussion on creation of two different rule bases in Section IV and we evaluate the rule-bases in Section V and conclude in Section VI.

## II. PREVIOUS WORK

The broad area of query intent identification is a well explored area in field of IR. While most query intent identification techniques are targeted towards the World Wide Web and Internet Search technology. Brenes [1] break up the problem of query intent identification into topic identification and intent identification. Broder [4] and Jansen [6] laid out the generally accepted taxonomy for intent identification by distinguishing between navigational and informational queries. Navigational queries on the web are those where the user intends to find a specific web page, whereas informational queries are when the user seeks information on a certain topic.

Our area of work, however, would fall under topic identification in the classification scheme proposed by Brenes [1]. Topic identification of queries is a promising area as it can dramatically improve accuracy of a retrieval system. Beitzel [2] compares query topic identification pre–retrieval with post-retrieval and observes the effect of training a classifier explicitly from pre topic tagged queries in contrast to mapping a classifier trained using a document classification scheme to a query topic classifier. In a previous work Beitzel [3] apply computational linguistics based technique to mine large collection of unlabeled web query logs that can create topical web query classifiers. They were able to demonstrate an improvement in topic identification and classification when their approach is combined with manual matching and supervised learning. Broader [5], Spink [11], and Pu [10] have all provided query taxonomies, similar to Yahoo! [13] and DMOZ [14] web directories. While topic identification of queries for the web is a fairly well explored topic, topic identification for mobile search has not been explored. Also most of these topic identification techniques are data driven and involve large training sets. In the absence of data a radically new approach is called for, that has minimal data dependency. This motivated us to develop a rule based system called Short Query Intent Identification System (SQIIS).

### III. OVERVIEW OF SQIIS

The approach taken in solving the topic intent identification problem involves very little lexical parsing. It is similar to the approach taken by Kopparapu et.all [7]. First, SQIIS accepts a short user query and identifies the intent of the query or the domain best matching the query. For example a query "*Chinese restaurant in Andheri*" would be detected as a search in the domain of yellow pages. Similarly a query "*Slumdog Millionaire in Andheri*" then would be identified as a movie domain query.

SQIIS caters to an SMS based querying system for a mobile search environment. Upon receiving the query, the system, makes a decision about which domain (topic) this query falls into by identifying the intent. The process that SQIIS uses for domain identification has two phases, namely, (1) Identifying word tokens using n-gram analysis of the input query and tagging the word tokens and (2) Rule-Based knowledge which uses the tagged tokens to determine the domain. SQIIS uses a rule-based $RB = \{r_1, r_2, \ldots, r_n\}$ which contains a set of rules. Each rule, $r_i = \{C_i, [d_1(c_1), d_2(c_2), \ldots d_n(c_n)]\}$ contains (1) a tag combination (a set of tags) $C_i = [t_1, t_3, \ldots, t_j]$ and (2) a set of domain $d_i(c_i)$ pairs where $d_i$ is the domain and $c_i$ is the confidence (or relevance) of that domain. The rule base is used to identify the domain and influences the performance of the system. Let $T = \{t_1, t_2, \ldots, t_n\}$ be the tag set $D = \{d_1, d_2, \ldots, d_s\}$ be the set of domains; one of which needs to be identified with a certain confidence given a tagged query. For example, if a query had two tags only, say t1 and t5, then the query would be represented as $\{1,0,0,0,1,0,,,,,0\}$; which would result in selection of d1 with $c_1$, $d_2$ with $c_2$ etc using the rule-base. The domain with the largest c would be the final selected domain due to the rule-base.

*A. Tagging*

SQIIS breaks the input query into word tokens through n gram analysis as a first step. Thus a token can contain one or more words in the input query. The process of tagging is performed on the tokens. Formally the process of tagging is performed as shown below.

*B. Algorithm for Tagging Word Tokens*

*Input:* The input query is a set of say j word tokens, namely, $W = \{w_1, w_3, \ldots, w_j\}$
*Output:* The output is a tag ($t_i \varepsilon T$) attached to each word token, namely, $Q = \{w_1[t_1], w_2[t_2], \ldots, w_j[t_j]\}$
*Procedure:* For each word token $w_1 \varepsilon W$
(1) Look up token in each of the lookup table in $\{l_1, l_2, ..l_i\}$ corresponding to each tag.
(2) If $w_i$ is found in a lookup table $l_j$ then attach the corresponding tag $t_j$ to the word token $w_i$.
(3) Each token is looked up in all the lookup tables. If the token returns a match in a specific lookup table then the tag associated with the lookup table is attached to that token. The output therefore would be of the form $Q = \{w_1$ [different list of tags attached], $w_2$ [different list of tags attached]......, $w_n$ [different list of tags attached].

*C. Rule Based Domain Identification Phase*

A rule based expert system approach is used for domain identification. Table 1 shows the data structure of the rule base. Each tuple (row) represents a rule. Each tuple has a combination of tags and a set of domains and confidences. If a particular combination of tags matches a tag combination associated with a rule, then the rule is "fired" and the domains and their corresponding confidences are returned. If a specific combination of tags occurs then the query can be said to belong to a specific domain with a certain degree of confidence. If the confidence is high for a certain domain for a specific combination of tags, then there is a higher probability that that query belongs or can be mapped accurately to that domain.

TABLE 1. RULE BASE DATA STRUCTURE

| Rule ID | Tag Combinations | Search Systems Domains | | |
|---|---|---|---|---|
| 1 | C1 => { $t_1$, $t_4$ } | D1(0.75) | D2(0.25) | …. |
| 2 | C2 => { $t_2$,$t_5$ } | D1(0.45) | D2(0.55) | …. |
| 3 | …. | …. | …. | …. |

We know that the output from the tagging phase is of the format $Q = \{w_1 [t_1], w_2 [t_2], \ldots, w_j [t_j]\}$. In this phase the query is transformed into a set of tags $Q_t = \{t_1, t_2, \ldots, t_j\}$. $Q_t$ is matched against the rule base (Table 1). If a combination in the rule base matches the set $Q_t$ then that rule (housed in the rule base) is fired and the domain confidences are returned. The domain with the maximum confidence is selected.

*D. Algorithm for Rule Based Domain Identification*

*Input:* $Q = \{w_1 [t_1], w_2 [t_2], \ldots, w_j [t_j]\}$
*Output:* Domain which has the highest confidence.
*Procedure:* Obtain $Q_t = \{t_1, t_3, t_4, t_j\}$ by and retaining the tags associated with the $w_j$ word tokens. Fit it to a combination $C_i$ in the rule table. Obtain domains and corresponding confidences.

*E. System Description*

The built functional system consists of three domains, Yellow Pages, Movie and Road Map domains. There are overall seven tags. Namely, T = {*address, category, proper name, movie title, movie performer (proper name), movie performer (category), direction word*}. The address tag is assigned to a generalized address element such as road, street or name of a road, street or city area. The category tag is a assigned to keywords such as *restaurant, theater, club, bank, hospital*. A proper name is a name of a person, business etc. The above described tags can be broadly associated with yellow pages. The direction word tag is associated with words like *from, to, way to* etc. The movie title proper name tag is associated with part or whole of movie names, movie performer category refers to keywords like actor, music director etc. The movie performer proper name refers to names of movie personalities. The tags mentioned are broadly associated with the movie domain. Ideally a certain combination of tags will lead to the identification of a domain.

## IV. RULE GENERATION STRATEGIES

Rule base construction is detrimental in the performance of the Short Query Intent Identification system. The rule-base is a mapping of combination of tags to all the domains with some weights. We used two different strategies to construct the rule-base. These are discussed next section.

*A. System Generated or Semi Supervised Rule Base*

The strategy for developing a rule base is a semi automated process. The method developed is based on the premise that if occurrence of a certain tag leads to a higher probability that the tag combination belongs to a particular domain. Here for each tag and domain pair, there is a weight attached, namely, $(t_i, d_j) \rightarrow w_{ij}$. When the tag $t_i$ occurs in the query then the weight $w_{ij}$ is attached to the domain j. Formally, for every tag $t_i$ in set T and domain $d_j$ in set D there is a weight $w_{ij}$ attached namely $(t_i, d_j) \rightarrow w_{ij}$. The confidence of a domain $d_k$, confidence($d_k$) for a tag combination $(t_1, t_2, \ldots, t_n)$, is given as confidence $(d_k) = \sum_{i=1}^{n} w_{ik}$. If there are a certain number of domains in the set D of domains within which the system operates, the confidences are normalized as norm-confidence $(d_k) =$

$$\frac{\sum_{i=1}^{n} w_{ij}}{\sum_{j=1}^{m} \text{confidence}(d_j)}$$

. This way for every combination of tags a set of confidences for each domain can be obtained. The domain with the maximum confidence is the domain to which the tag combination and its associated query are classified. Also some tags can not co-occur with other tags. From example a "category" tag might not exist with a "movie title proper name" tag. So we construct a set of tag pairs $(t_i, t_j)$ where the tags can not appear together in a combination. If the tags appear together in a combination then the combination is considered invalid. In a very loose sense tag dimensionality checking is done automatically.

The system generated rule base is easier to construct as it required determining a gross set of rules. It is less prone to human error, but chances of inaccuracies are more because it does not analyze the rule construction at a deep level.

*B. Hand Crafted Rule Base*

In a hand crafted rule-base say there are n tags and m domains associated with the system. We enumerate all the possible tag occurrences, namely, $2^n$ combinations. Manually we go through each combination and then determine the combination best describes a particular domain. Here we can say with absolute certainty if a combination can be classified in a particular domain. The domain confidence would be 1 and the other domain confidences would be 0. However some combinations might not be classifiable in any domain. Under those circumstances we would say that no domain can be determined. The system using hand crafted rule-base should perform best. However, hand-crafting needs tedious human intervention and is prone to human errors.

In hand crafting the domain confidences are usually absolute. For example if a combination returns a domain $d_i$, then we say that the confidence of $d_i$ is 1 while all other domains have a confidence of 0. In system generated rule-bases, domain confidences are not absolute. For each domain the domain confidence may vary between 0 and 1.

## V. EXPERIMENT AND EVALUATION

We evaluated the two rule-bases we constructed for our system. In our current system we had 7 tags and 3 domains. We generated using a software program all possible combinations namely $2^7 - 1$ or 127 tag combinations. We treated these combinations as output from the tagging stage and fed them into our rule base. We evaluated all tag combinations of size between 1 and 6 which resulted in a certain tag combinations. Each of these combinations is fed into the rule-base. If the tag combination returns a specific domain then error is introduced with one tag at a time being changed to a different tag. Thus erroneous tag combinations are obtained and fed into the same rule-base and in each case the change in output is noted. If there is no change in outcome (final domain remains same, even after error in tagging) the case is counted as a C0. If there is a change in the outcome (final domain changes) then the case is classified as a C1. If the alteration results in a system being unable to detect topical intent, we count it as a C2. We also computed the change in the identified domain confidences between the actual output and the desired output as a Euclidean distance between the domain confidences of the desired output and input. Let the output before error in tagging be $[d_1(c_1), d_2(c_2),\ldots d_m(c_m)]$ assuming m domains in the system. Here $c_1, c_2 \ldots c_m$ are the confidences for domains $d_1, d_d$ and $d_m$ respectively. In our case m = 3. Let the output due to error in tagging be $[d_1(c'_1), d_2(c'_2), \ldots d_m(c'_m)]$. Then the distance is computed as $\sqrt{\sum_{i=1}^{m}(c_i - c'_i)^2}$. The results for each of the two cases, namely C0 and C1 for the system generated and hand crafted rule base and rule base are shown in Figures 1, and 2 respectively. Figures 1 and 2 show cumulative plots of the variation from the desired output for misclassification for system generated rules (y-axis) versus the number of cases (x-axis). The results for handcrafted rules are tabulated in Table 2 and 3.

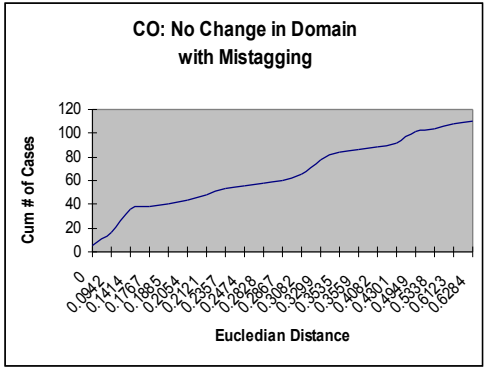

Figure 1.  Results for System Generated Rule for cases in C0

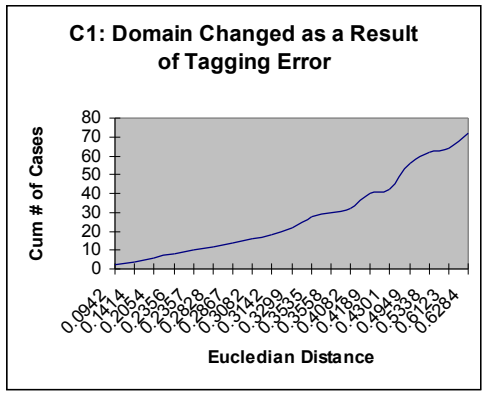

Figure 2.  Results for System Generated Rule for cases in C1

TABLE 2: RESULTS FOR C0 FOR HAND CRAFTED RULES

| Euclidian Distance | # of Cases | Cum # of Cases |
|---|---|---|
| 0 | 106 | 106 |

TABLE 3: RESULTS FOR C1 FOR HAND CRAFTED RULES

| Euclidian Distance | # of Cases | Cum # of Cases |
|---|---|---|
| 1.4142 | 92 | 92 |

TABLE 4: RESULTS FOR C0 FOR SYSTEM GENERATED RULES

| Euclidian Distance | # of Cases | Cum # of Cases |
|---|---|---|
| 0 | 6 | 6 |
| 0.0942 | 10 | 16 |
| 0.1414 | 20 | 36 |
| 0.1767 | 2 | 38 |
| 0.1885 | 2 | 40 |
| 0.2054 | 4 | 44 |
| 0.2121 | 4 | 48 |
| 0.2357 | 6 | 54 |
| 0.2474 | 2 | 56 |
| 0.2828 | 2 | 58 |
| 0.2867 | 2 | 60 |
| 0.3082 | 6 | 66 |
| 0.3299 | 12 | 78 |
| 0.3535 | 6 | 84 |
| 0.3559 | 2 | 86 |
| 0.4082 | 2 | 88 |
| 0.4301 | 4 | 92 |
| 0.4949 | 10 | 102 |
| 0.5338 | 2 | 104 |
| 0.6123 | 4 | 108 |
| 0.6284 | 2 | 110 |

Table 4, 5 show tabulated results for C0, C1 for System Generated Rules. A quick glance through the results tabulated in Tables 2 through 5, might give the impression that the system generated rule is performing better than the hand crafted rule. However, notice that the variation measured in Euclidian distance is the least (0) in the hand crafted rules. Assuming a reasonable threshold $\tau = 0.6$ it can be seen that for C0 cases, the cumulative number of cases for the hand-crafted set = 106. However for the same threshold the system generated rule set is 110. We can conclude that the hand crafted rule actually performs a lot better (as expected).

TABLE 5: RESULTS FOR C1 FOR SYSTEM GENERATED RULES

| Euclidian Distance | # of Cases | Cum # of Cases |
|---|---|---|
| 0.0942 | 2 | 2 |
| 0.1414 | 2 | 4 |
| 0.2054 | 2 | 6 |
| 0.2356 | 2 | 8 |
| 0.2357 | 2 | 10 |
| 0.2828 | 2 | 12 |
| 0.2867 | 2 | 14 |
| 0.3082 | 2 | 16 |
| 0.3142 | 2 | 18 |
| 0.3299 | 4 | 22 |
| 0.3535 | 6 | 28 |
| 0.3558 | 2 | 30 |
| 0.4082 | 2 | 32 |
| 0.4189 | 8 | 40 |
| 0.4301 | 2 | 42 |
| 0.4949 | 14 | 56 |
| 0.5338 | 6 | 62 |
| 0.6123 | 2 | 64 |
| 0.6284 | 8 | 72 |

## VI. CONCLUSIONS AND FUTURE WORK

In this paper we have described the functioning of a rule based Short Query Intent Identification System. The rule base is a critical component of SQIIS. We have described two techniques (system generated and hand crafted) for generating a rule base. We have demonstrated how the hand crafted rule base performs better than the system generated rule base. Our experiments involved three domains yellow pages, movies and road map navigation domains. For domain detection we created a methodology for identifying named entities and tagging them with custom tags. In our system we have currently 7 tags. As future work we would be expanding this system to detect query topics spanning over a wide spectrum of domains and improving rule based generation strategies.